\documentclass[twocolumn,showpacs,amsmath,amssymb,prc,groupedaddress]{revtex4-1}
\usepackage[dvipdfm,colorlinks]{hyperref}
\usepackage{graphicx,epsfig,amsmath,graphicx,amssymb,wasysym}
\usepackage[english]{babel}
\usepackage{t1enc}
\usepackage[latin2]{inputenc}
\begin{document}
\title{Multipole solution of hydrodynamics and higher order harmonics}

\author{M.~Csan\'ad}
\email{csanad@elte.hu}
\homepage{http://csanad.web.elte.hu/}
\author{A. Szab\'o}
\affiliation{E\"otv\"os University, Department of Atomic Physics, Pázmány P. s. 1/a, H-1117 Budapest, Hungary}

\date{\today}

\begin{abstract}
The time evolution of the medium created in heavy ion collisions can be described by hydrodynamical models.
After expansion and cooling, the hadrons are created in a freeze-out. Their distribution describes the final state of
this medium. In particular their azimuthal asymmetry, characterized by the elliptic flow coefficient $v_2$, is
one of the most important observables in heavy ion physics. In recent years it has been revealed that if
measuring relative to higher order event planes $\Psi_n$, higher order flow coefficients $v_n$ for $n>2$ can be
measured. This is due to initial state fluctuations, previously not described by analytic solutions of relativistic
hydrodynamics. In this paper we show the first solutions that utilize higher order asymmetries and thus
yield realistic $v_n$ flow coefficients. It is a clear consequence of this that different flow patterns may
lead to the same observed flow coefficients. We also compare our results to PHENIX measurements and
determine a possible parameter set corresponding to these data.
\end{abstract}
\maketitle

\section{Introduction}
It is well known that the medium created in high energy heavy ion collisions can be described with perfect fluid
hydrodynamics; in particular the soft hadron production can be successfully compared to hydrodynamic models~\cite{Adcox:2004mh}.
Exact solutions provide an analytic handle on the connection between the initial state, the
dynamic parameters of the system and the observables. Usually elliptical symmetry is assumed in the transverse
plane~\cite{Csorgo:2003ry}, as this is simple to handle and represents geometries that yield realistic results
for spectra, Bose-Einstein correlation functions and elliptic flow. However, nuclei contain a finite number of nucleons,
are thus not exactly spherically symmetric, and their overlap region also fluctuates on an event-by-event basis.
This results in an event-by-event fluctuating initial condition, and gives rise to nonzero high order flow coefficients,
with respect to higher order reaction planes~\cite{Adare:2011tg,ALICE:2011ab,Adamczyk:2013waa}. This
was successfully reproduced in numerical hydrodynamical calculations in Refs.~\cite{Petersen:2010cw,Schenke:2010rr} abd
more recently in Refs.~\cite{Bravina:2013xla,Chatterjee:2014nta}.

In this work we show the first exact analytic solutions of relativistic hydrodynamics that assume higher order asymmetries.
An important point of our work is giving explicit examples of analytic flow patterns leading to realistic higher order flow coefficients.
Our paper is organized as follows. First we give a short introduction to relativistic perfect fluid hydrodynamics. In the following
section we describe our new solution, its properties and its relations to known solutions. Then we present model results on
observables, such as transverse momentum spectra and angular anisotropy coefficients (or harmonics) $v_n$.
Finally we show a comparison of our results to PHENIX measurements of Ref.~\cite{Adare:2011tg}.

\section{Perfect fluid hydrodynamics}
In this manuscript we adopt the following notation: $\varepsilon$ is energy density, $p$ is pressure, $n$ (if present) is 
the density of a conserved charge and $\sigma$ is entropy density. Moreover, $g^{\mu\nu}$ is
the metric tensor, diag$(-1,1,1,1)$, while $x^{\mu} = (t, r_x, r_y, r_z)$ is a given point in space-time
(sometimes, for the sake of simplicity, denoted by $x$ without superscript),
$\tau = \sqrt{t^2-r^2}$ is the coordinate proper time, $\partial_{\mu} = \frac{\partial}{\partial x^{\mu}}$ is the
derivative versus space-time, while  $p^{\mu} = (E, p_x, p_y, p_z)$ is the four-momentum (also sometimes
denoted by $p$ without superscript). The equations of hydrodynamics then are
\begin{align}
\partial_\mu (n u^\mu) & = 0,\\
\partial_\nu T^{\mu \nu} & = 0.
\end{align}
The fluid is perfect if the energy-momentum tensor $T^{\mu\nu}$ is diagonal in the local rest frame, i.e.,  viscosity and heat conduction are
negligible. This can be assured if $T^{\mu\nu}$ is chosen as
\begin{align}
T^{\mu\nu}=(\varepsilon+p)u^\mu u^\nu-pg^{\mu\nu}.
\end{align}
If there are no conserved charges in this perfect fluid, an other local conservation equation may be written: that of
entropy density $\sigma$.

An analytic hydrodynamical solution is a functional form for $u^\mu$, $\varepsilon$, $p$ and $n$ or $\sigma$, which solves
the above equations. These quantities are also subject to the equation of state (EoS), which closes the set of equations. 
Usually $\varepsilon = \kappa p$ is chosen, where $\kappa$ may depend on temperature $T$, and solutions with temperature
dependent $\kappa$ were found in Ref.~\cite{Csanad:2012hr}. 
In this paper, however, we use a solution with constant $\kappa$. It is
important to see that in this case $\kappa = 1/c_s^2$, with $c_s$ being the speed of sound.
Temperature can then be defined based on entropy density, energy density, and pressure.
An important result for hydrodynamic models is that, because hadrons are created at the quark-hadron transition,
hadronic observables do not depend on the initial state or the dynamical equations (equation of state) separately,
just through the final state~\cite{Csanad:2009sk}.

There was a long search for exact solutions of relativistic hydrodynamics, and only a few applicable ones were found.
The first exact solutions of relativistic hydrodynamics were described by Landau and collaborators,
calculating momentum distributions of produced particles in high energy collisions from the
theory of locally thermalized and relativistically expanding fluids~\cite{Landau:1953gs,Belenkij:1956cd}. 
These solutions were given in an implicit form. The first exact and explicit solutions were found by Hwa~\cite{Hwa:1974gn}
and later, independently by Bjorken~\cite{Bjorken:1982qr}. A unified description of these models was found in Ref.~\cite{Bialas:2007iu} 
and extended to general 1+1 dimensional relativistic flows in Refs.~\cite{Beuf:2008vd,Peschanski:2010cs}.
Another one-dimensional (1D), longitudinally expanding explicit relativistic solution has been
found in Ref.~\cite{Csorgo:2002ki}, generalized later to axial symmetry
in 3D~\cite{Csorgo:2003rt} and ellipsoidal symmetry as well~\cite{Csorgo:2003ry}.
Other multi-dimensional solutions of relativistic hydrodynamics were given in
Refs.~\cite{Liao:2009zg,Lin:2009kv,Gubser:2010ze,Hatta:2014gqa}.
However, a realistic elliptic flow could not have been calculated from most of these models (except the one
mentioned in Ref.~\cite{Csanad:2009wc}), let alone higher order azimuthal asymmetries. In the next section we report the first
exact analytic solution of relativistic hydrodynamics that yields higher order asymmetries with measurements.

\section{Multipole solutions}
Let us start from the solution given in Ref.~\cite{Csorgo:2003ry}, a (1+3)D relativistic solution with realistic
(not spherically symmetric) geometry. Here the thermodynamical quantities are
(at a given proper time) constant on the surfaces of an expanding ellipsoid, defined by the $s$ scale variable,
\begin{align}
s=\frac{r_x^2}{X^2}+\frac{r_y^2}{Y^2}+\frac{r_z^2}{Z^2},\label{e:scale2}
\end{align}
where $r_x$,$r_y$,$r_z$ are the spatial coordinates, while $X$, $Y$, $Z$ are the time dependent axes of the ellipsoid.
The velocity profile is given as
\begin{align}
u^{\mu}=\gamma\left( 1,\frac{\dot{X}}{X} r_x,\frac{\dot{Y}}{Y} r_y,\frac{\dot{Z}}{Z} r_z \right),
\end{align}
where $\dot{X}=dX/dt$ and similarly for $Y$ and $Z$. 

If the comoving derivative of $s$ vanishes, i.e., $u^\mu \partial_\mu s=0$, then we can construct a hydrodynamical
solution with $s$ being its scaling parameter. For the above equation to be fulfilled, we need $\dot{X}$,$\dot{Y}$,$\dot{Z}=$ constant.
If we choose $X=\dot{X} t$, $Y=\dot{Y} t$, $Z=\dot{Z} t$, then with $\tau=\sqrt{x_{\mu}x^{\mu}}$ we get
\begin{align}
u^{\mu}=\frac{x^{\mu}}{\tau},
\end{align}
with $x^{\mu}$ being the space-time coordinates and $\tau$ the coordinate proper time.

The thermodynamic quantities are then given with an arbitrary $\nu(s)$ scale function as:
\begin{align}
n(x)&=n_f \left(\frac{\tau_f}{\tau}\right)^3 \nu(s),\\
T(x)&=T_f\left(\frac{\tau_f}{\tau}\right)^{3/{\kappa}}\frac{1}{\nu(s)},\\
p(x)&=p_f\left(\frac{\tau_f}{\tau}\right)^{3+3/{\kappa}},
\end{align}
where $n(x)$ is the number-density of a conserved charge (if any), $T(x)$ is temperature, $p(x)$ is pressure, and constants are
normalized via $p_f=n_fT_f$. Parameters with the index $f$ are values of the given quantity at the freeze-out (and if
the quantity has also spatial dependence, then in the center of the fireball), in particular $\tau_f$ is the freeze-out
proper time, when hadronization occurs. Note that this solution (and any other of $\kappa=$ const. type) can be written
up for the entropy density $\sigma(x)$ instead of $n(x)$ identically~\cite{Csanad:2012hr}. This means that here
$\sigma(x) = \sigma_f \left(\tau_f/\tau\right)^3 \nu(s)$ can be taken, and $n(x)$ shall not be used,
if there are no conserved charges in the system.

\begin{figure}
\centering
\includegraphics[angle=270,width=0.5\textwidth]{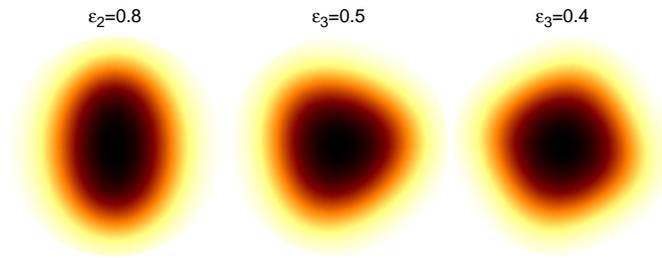}
\caption{(Color online) Heat map of $s$ values in the transverse plane for the $N=2,3,4$ solutions, respectively. If the temperature
is a monotonic continuous function of $s$, then this is homomorphic with the actual temperature distribution of the solution.}
\label{f:s234}
\end{figure}

Let us now show how the above known solution can be extended to multipole symmetries. First, let us consider a
1+2 dimensional case. If we rewrite the scale variable $s$ [given in Eq.~(\ref{e:scale2})] to polar coordinates
(with $x=r\sin\phi$, $y=r\cos\phi$) we get
\begin{align}
s&=\frac{r^2}{R^2}\left(1+\epsilon \cos(2\phi)\right)\textnormal{, where}\\
\frac{1}{R^2}&=\frac{1}{X^2}+\frac{1}{Y^2}\textnormal{ and }\epsilon(t) = \frac{X^2+Y^2}{X^2-Y^2},\\
\end{align}
i.e., $R$ is the average system size and $\epsilon$ the eccentricity.  As $X$ and $Y$ are time dependent, $\epsilon$ may
also depend on time. However, if $X$ and $Y$ are both proportional to time, this dependence cancels and
$\epsilon(t)=\epsilon$ remains constant. The above formula for $s$ can be generalized
to higher order symmetries:
\begin{align}
s=\frac{r^N}{R^N}\left(1+\epsilon(t) \cos(N\phi)\right)\label{e:sN}
\end{align}
where $N$ is the order of the symmetry. To visualize this, we show a heat map of $s$ values for several different $N$
values in Fig.~\ref{f:s234}. 

With the $s$ given in Eq.~(\ref{e:sN}), we can derive a new solution:
\begin{align}
u^{\mu}(x)&=\gamma \left(1, \frac{\dot{R}}{R(t)}r\cos\phi,\frac{\dot{R}}{R(t)}r\sin\phi\right),\\
n(x)&=n_f\left( \frac{\gamma R_f}{R(t)} \right)^2 \nu(s),\\
T(x)&=T_f\left(\frac{\gamma R_f}{R(t)}\right)^{2/{\kappa}}\frac{1}{\nu(s)},\\
p(x)&=p_f\left( \frac{\gamma R_f}{R(t)} \right)^{2+2/\kappa},
\end{align}
where $\tau_f$ is again freeze-out proper time, $R(t)=u_t t$ (i.e., $\dot{R}=u_t=$ const. being the expansion velocity),
$R_f=u_t\tau_f$ , $\gamma=\frac{1}{\sqrt{1-r^2 \dot{R}^2/R^2}}$, and $u_t$ is a
transverse expansion velocity, while $\epsilon=$ constant. In this case we obtain a Hubble-flow profile, as
\begin{align}
\frac{\gamma R_f}{R(t)} = \frac{\tau_f}{\tau},
\end{align}
and $u^\mu=x^\mu/\tau$.
Note again, that instead of $n(x)$, $\sigma(x)$ can be written up the same way, if there are no conserved charges.ó

This solution can be generalized to 1+3 dimensions multiple ways. We may choose cylindrical coordinates $(r,\phi,z)$, and add
a $z^N/R^N$ term to $s$:
\begin{align}
s&=\frac{r^N}{R^N}\left(1+\epsilon \cos(N\phi)\right)+\frac{z^N}{R^N},\\
u^{\mu}(x)&=\frac{x^\mu}{\tau},\\
n(x)&=n_f\left( \frac{\tau_f}{\tau} \right)^3 \nu(s),\\
T(x)&=T_f\left(\frac{\tau_f}{\tau} \right)^{3/\kappa}\frac{1}{\nu(s)},\\
p(x)&=p_f\left( \frac{\tau_f}{\tau} \right)^{3+3/\kappa}.
\end{align}

We get another solution in spherical coordinates if we write $s$ as
\begin{align}
s=\frac{r^N}{R^N}\left\{1+\epsilon_a \cos(N\phi)[1-\cos(N\theta)]+\epsilon_b \cos(N\theta) \right\}\label{e:sN3D}
\end{align}
where $\epsilon_a$ and $\epsilon_b$ are eccentricities in different planes. There are many other type
of scale variables possible, and it turns out that there is a relatively high level of freedom in the choice of scale
variables, as it was already mentioned in Ref.~\cite{Csorgo:2003ry}. They indicate that any $F(r_x^2/t^2,r_y^2/t^2,r_z^2/t^2)$
function provides a valid scaling variable. Our solution falls in a somewhat more general class, where the scaling variable is
given as $s=F(r_x/t,r_y/t,r_z/t)$, with an arbitrary $F$ function works [the square has to be dropped, as in case of odd $N$'s,
$\cos(N\phi)$ is not a function of $r_i^2/t^2$ but of $r_i/t$].

\begin{figure}
\centering
\includegraphics[angle=270,width=0.5\textwidth]{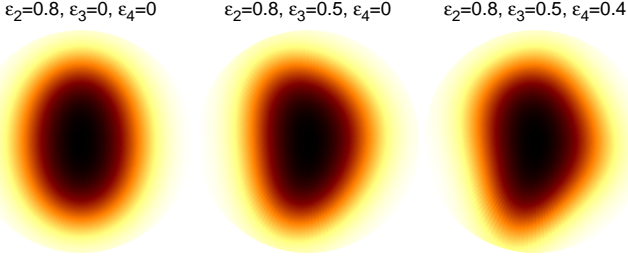}
\caption{(Color online) Heat map of $s$ values in the transverse plane, with multiple superimposed symmetries. The more $\epsilon_N$
components are included, the more asymmetric the shape gets.}
\label{f:multis}
\end{figure}

We may also combine several symmetries with different $N$'s via 
\begin{align}
s=\sum\limits_N \frac{r^N}{R^N}\left\{1+\epsilon_N\cos[N(\phi-\psi_N)]\right\}
\end{align}
with $\psi_N$ being the $N$th order reaction planes (which cancel from the observables). This way we get
new solutions with almost arbitrary shaped initial distributions, see Fig.~\ref{f:multis}. It is important to 
note here that although the initial state fluctuation in the observed collision is present through the
orientation of the $N$th order reaction planes and the strength of higher order asymmetries, the
event plane orientation itself does not affect the measured quantities. Thus if every $v_N$ is
measured relative to the $N$th order reaction plane, then the (event-through-event) averaged
value of $v_N$ will correspond to an average $n$-pole anisotropy $\epsilon_N$. Note also that
our solution, presented above, contains flow patterns belonging to a special class of initial conditions,
defined by the energy density profile and Hubble flow. In a realistic scenario, initial conditions contain
more sophisticated inhomogeneities in the density distributions, and velocity distributions are also more
complicated. Our paper's goal is, however, to explicitly show flow patterns (exact hydro solutions) that
describe multipole expansions and lead to realistic observable flow asymmetries. To arrive at this goal,
let us calculate observables from our solutions.

\begin{table}
\caption{Typical values and meaning of model parameters. Values were partly taken from Ref.~\cite{Csanad:2009wc}.
\label{tab:params}}
\centering
\begin{tabular}{|c|c|c|}
\hline
variable&            typical value&	meaning\\
\hline
$T_f$&              $200$ MeV &	central freeze-out temperature\\
$u_t$&              $0.6$&	transverse expansion\\   
$b$&		     $0.08$&	$\sim$ temperature gradient\\
$\tau_f$ &         $7.7$ fm$/c$&	freeze-out proper time\\
$\epsilon_2$&   $0.50$&	elliptic eccentricity\\
$\epsilon_3$&   $0.25$&	triangular eccentricity\\
$\epsilon_4$&   $0.08$&	quadrupole eccentricity\\
\hline
\end{tabular}
\end{table}

\section{Observables}
Similarly to Ref.~\cite{Csanad:2009wc}, we use a freeze-out (FO) scenario in which the pre FO medium is described by 
hydrodynamics, and the post FO medium is that of observed hadrons.  In our framework we assume
that the freeze-out happens at a given proper time, e.g., due to a self-quenching effect or if the
phase space evolution is that of a collisionless gas. Thus there is no jump in the equation of state post-
and pre-FO, i.e., $\kappa$ goes to $\kappa_{\rm free}$ smoothly, to the EoS of free hadrons. 
In this case the hadronic observables can be extracted from the solution via the phase-space distribution at
the FO. This will correspond to the hadronic final state or source distribution $S(x, p)$.
We also do not need to fix a special equation of state, because the same final state can be achieved with 
different equations of state or initial conditions~\cite{Csanad:2009sk}. Thus in this paper $\kappa$ is arbitrary
-- the hadronic observables do not restrict its value. Based on the above, the source distribution takes the following
form:
\begin{align}
S(x,p)d^4x=\frac{g}{(2\pi)^3} n(x)e^{-p_{\mu}u^{\mu}(x)/T(x)}  H(\tau) p_{\mu} d^3\Sigma_{\mu}(x) dt
\end{align}
where $g$ is the degeneracy factor of the given particle species, $H(\tau)$ is the proper-time probability distribution
of the FO (assumed to be a delta distribution), the exponential with the temperature stems from
the Boltzmann--J\''uttner-distribution, and $d^3\Sigma_{\mu}(x)$
is the vector measure of the freeze-out hypersurface (which gives the Cooper-Frye flux factor, if multiplied by $p^\mu$).
If the freeze-out is a delta distribution at a given $\tau$, this vector measure can be given as $\frac{u^{\mu}d^3x}{u^0}$.
Finally, our distribution is
\begin{align}
S(x,p)d^4x=\frac{g}{(2\pi)^3}n(x)e^{-p_{\mu}u^{\mu}(x)/T(x)} \delta(\tau-\tau_f) \frac{p_{\mu}u^{\mu}}{u^0}d^4x,
\end{align}
where $T(x)$, $u^\mu(x)$, and $n(x)$ are defined by the hydrodynamic solution. From this, observables can be 
calculated via integrals:
\begin{align}
N_1(\mathbf{p})&=E\frac{d^3n}{d^3\mathbf{p}}=\int S(x,\mathbf{p})d^4x\label{e:Npalpha},\\
N_1(p_t)&=\left.\frac{dn}{2\pi p_tdp_t}\right|_{y=0}=\frac{1}{2\pi}\int\limits_0^{2\pi} \left.N(\mathbf{p})\right|_{p_z=0} d\alpha,\label{e:Np}
\end{align}
where $\mathbf{p}=(p_t\sin\alpha,p_t\cos\alpha,p_z)$ is the three-dimensional momentum, $p_z$ its longitudinal
and $p_t$ its transverse component, while $\alpha$ is its angle in the transverse plane. We restrict ourselves to
midrapidity observables, so we use $p_z=0$ (or rapidity $y=0$), and define transverse momentum flow coefficients as follows:
\begin{align}
v_n(p_t)=\frac{\frac{1}{2\pi}\int\limits_0^{2\pi}\left.N(\mathbf{p})\right|_{p_z=0}\cos(n\alpha)d\alpha}{N_1(p_t)}
=\langle\cos(n\alpha)\rangle \label{e:vn}.
\end{align}

Let us now calculate the integral of Eq.~(\ref{e:Npalpha}). If we choose a scale function of exponential form, $\exp (-bs)$
(i.e., the fireball is the hottest in the center and has a spatially Gaussian profile) we get
\begin{align}
N_1(p_t)\propto &\int \nu(s)
\exp\left[\frac{p_t\cos(\alpha-\phi)-Et}{\tau T_f}\nu(s)\left(\frac{\tau_f}{\tau}\right)^{-\frac{3}{\kappa}}\right]\nonumber\\
\times&\delta(\tau-\tau_f)\frac{\tau}{t}\frac{Et-rp_t\cos(\alpha-\phi)}{\tau}d^4x d\alpha
\end{align}
Now let us make an integral transformation from $t$ to $\tau$; then the result is
\begin{align}
&N_1(p_t)\propto\int e^{bs}\exp\left[\frac{rp_t\cos(\alpha-\phi)-E\sqrt{\tau_f^2+r^2+z^2}}{\tau_f T_f e^{-bs}} \right]
\nonumber\\
&\times\frac{E\sqrt{\tau_f^2+r^2+z^2}-rp_t\cos(\alpha-\phi)}{\tau_f^2+r^2+z^2}r\tau_fdrd\phi dzd\alpha
\end{align}
Values for $v_n(p_t)$ can be calculated similarly, as defined in Eq.~(\ref{e:vn}).
 
\begin{figure}
\centering
\includegraphics[angle=270,width=0.7\linewidth]{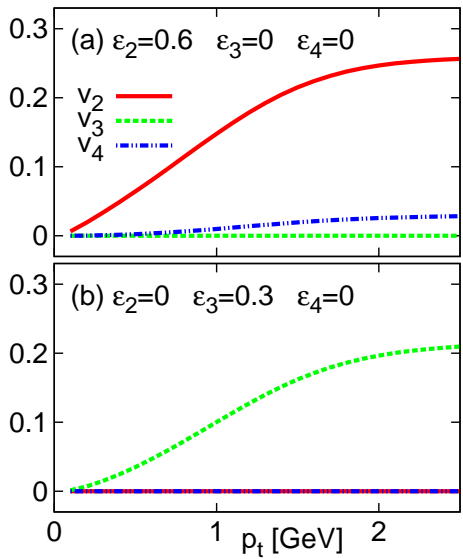}
\caption{(Color online) Curves for $v_2$,$v_3$,$v_4$ with only $\epsilon_2\neq 0$ are shown in the top panel (a), while for 
 $\epsilon_3 \neq 0$ only are shown in the bottom panel (b). Clearly there is no ``interference'' between odd and even  harmonics.}
\label{fig:epstest}
\end{figure}

Let us analyze the results from this model. Parameters other than higher order anisotropies ($\epsilon_n$)
can be taken from Ref.~\cite{Csanad:2009wc}, as summarized in Table~\ref{tab:params}. 
Note that azimuthally integrated observables are not sensitive to the anisotropies of this model, so
spectra and HBT with parameters from Table~\ref{tab:params} are compatible with PHENIX 200 GeV
Au+Au data, as results from this model are the same as from those in Ref.~\cite{Csanad:2009wc}.
We calculated $v_n$ for $n=2,3,4$ with only one $\epsilon_n \neq 0$. Clearly the odd and even
harmonics don not ``mix''; i.e., if only $\epsilon_3 \neq 0$ then only $v_3\neq 0$, however, $\epsilon_2$
gives rise to a nonzero $v_2$ and $v_4$. See results in Fig.~\ref{fig:epstest}.

\begin{figure*}
\centering
\includegraphics[angle=270,width=1\textwidth]{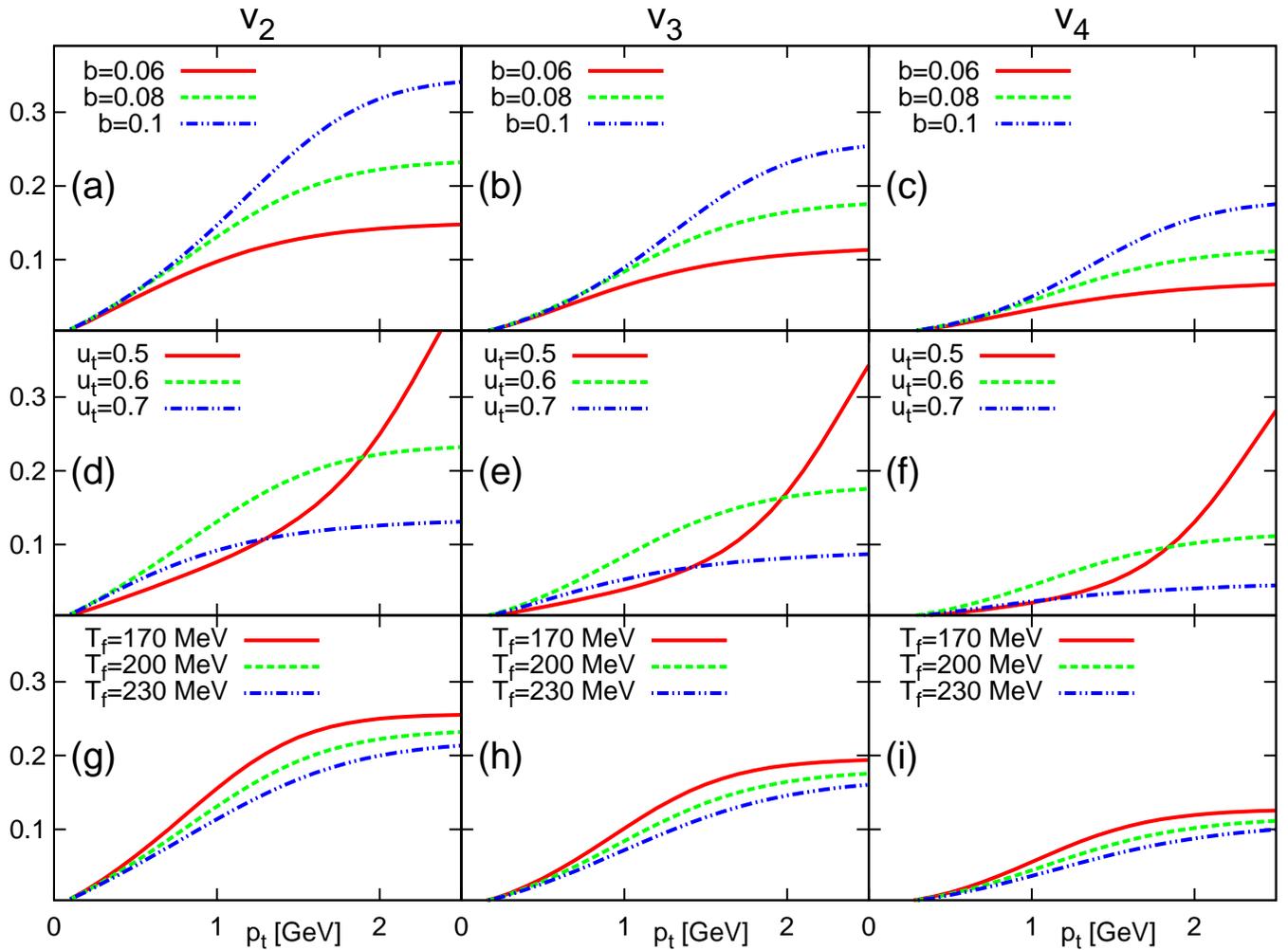}
\caption{(Color online) The $v_2$,$v_3$,$v_4$ coefficients at different values of $b$ [panels (a)-(c)], $T_f$ [panels (d)-(f)] and $u_t$ [panels (g)-(i)]. The other, fixed parameters were taken from Table~\ref{tab:params}.}
\label{fig:pardeps}
\end{figure*}

In Fig.~\ref{fig:pardeps} we investigate the parameter dependence of the results of this
model for $v_n(p_t)$. We vary one parameter, and fix the rest to values from Table~\ref{tab:params}.
In this model, $u_t$, and $b$ have a strong effect on the $v_n$ coefficients. In the Ref.~\cite{Csanad:2009wc},
model results only depend on $u_t^2/b$, but with the scale variable $s$ used here, terms in $s$ depend
on $u_t^N/b$ factors for various $N$ values. Thus the $v_n$ parameters depend here on both $b$ and $u_t$.
This dependence is, however, strongly coupled, as we will see later on.

\section{Data comparison}

\begin{table}
\caption{Model parameters (with statistical errors) from data fit  to PHENIX 200 GeV Au+Au data~\cite{Adare:2011tg}.
Parameter $b$
(governing the temperature gradient) is strongly correlated to the other parameters, so only a confidence interval
could have been given for it. However, it affects the value of the other parameters, and this results in a systematic
uncertainty of them. This is around 17\% for $u_t$, 27\% for $\epsilon_2$, 8\% for $\epsilon_3$ and
9\% for $\epsilon_4$ (independently of centrality). The magnitude of this systematic error is visualized in Fig.~\ref{fig:pars}.
\label{tab:fitpars}}
\centering
\begin{tabular}{|c|c|c|c|c|c|c|}
\hline
 &0-10\%&10-20\%&20-30\%&30-40\%&40-50\%\\
\hline
$u_t$ [\permil]& 740$\pm$3 & 765$\pm$2 & 781$\pm$2 & 787$\pm$2 & 774$\pm$3\\
$\epsilon_2$ [\permil]& 175$\pm$2 & 330$\pm$2 & 473$\pm$3 & 571$\pm$4 & 621$\pm$6\\
$\epsilon_3$ [\permil]& 99$\pm$2 & 136$\pm$2 & 165$\pm$2 & 180$\pm$3 & 182$\pm$4\\
$\epsilon_4$ [\permil]& 44$\pm$2 & 69$\pm$2 & 96$\pm$3 & 111$\pm$5 & 125$\pm$12\\\hline
$b$                & \multicolumn{5}{c|}{0.05$-$0.2}\\
\hline
\end{tabular}
\end{table}

\begin{figure*}
\centering
\includegraphics[angle=270,width=0.9\textwidth]{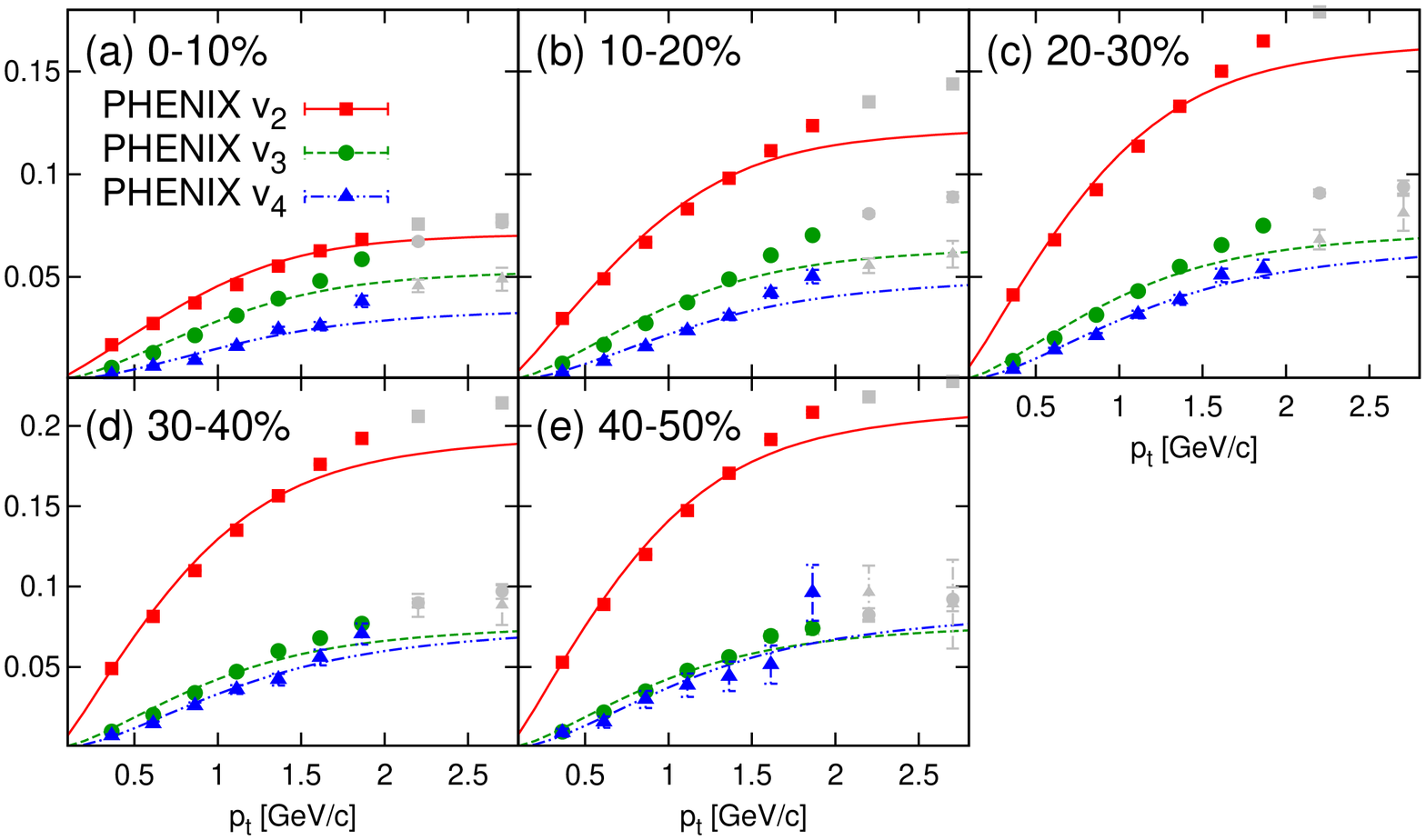}
\caption{(Color online) Fits to PHENIX 200 GeV Au+Au data~\cite{Adare:2011tg} in five centrality bins. Fit parameters are summarized
in Table~\ref{tab:fitpars}. It is important to note that many set of parameters (and thus many different flow patterns) lead to the
same observed $v_n$ values, and a large set of observables (or constraints on the initial conditions, as done for numerical
calculations) are needed to determine model parameters.}
\label{fig:fits}
\end{figure*}

In this section we compare our results to PHENIX data on higher order harmonics measured in 200 GeV Au+Au
collisions~\cite{Adare:2011tg}. Fit parameters of the model are $\epsilon_N$ (for $N=2,3,4$), $u_t$ and $b$
($T_f$ and $\tau_f$ were fixed to values given from spectra and HBT comparisons of a similar model, described
in Ref.~\cite{Csanad:2009wc}). However, there was a strong correlation between $b$ and the other parameters.
We scanned the parameter space for lowest $\chi^2$ values, but found only a weak dependence on $b$ itself:
this parameter yielded approximately the same curve for $b\in[0.05,0.2]$, and this resulted in a systematic error
for the other parameters coming from uncertainty of the $b$ value. This explicitly shows that different flow patterns
(different parameters of our solution) may lead to the same observables.
Model fits are shown in Fig.~\ref{fig:fits}.
Around $p_t=2$ GeV, non-hydro effects start to play an important role, thus we did not fit data points above
this value. Model parameters from the fit are summarized in Table~\ref{tab:fitpars}. It is important to note
that even though higher order flow coefficients arise from event-by-event fluctuations, an average triangular
or quadrupole anisotropy can be extracted from the data this way. This extraction is somewhat ambiguous however,
due to the correlation of $b$ and $u_t$ parameters. Also note that we did not vary parameters that were fixed
based on Ref.~\cite{Csanad:2009wc} -- these would introduce even more ambiguity, and more data are needed
to fix them (as done in~\cite{Csanad:2009wc}).

\begin{figure*}
\centering
\includegraphics[angle=270,width=0.6\textwidth]{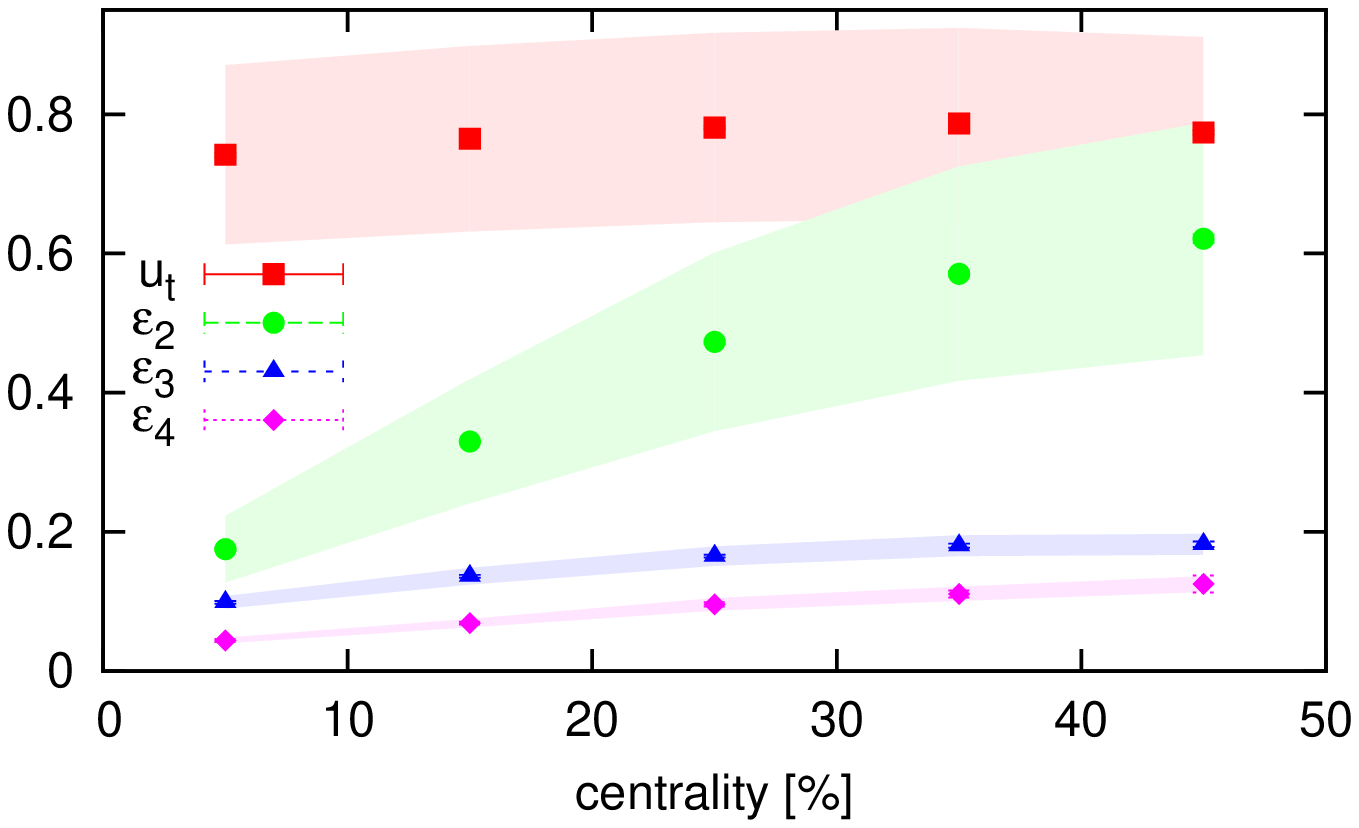}
\caption{(Color online) Model parameters from data fit to PHENIX 200 GeV Au+Au data~\cite{Adare:2011tg}, as a function
of centrality. Systematic error band comes from the correlation with $b$, see caption of Table~\ref{tab:fitpars}. Note that if $b$
depends on centrality (which is a realistic scenario) then of course $u_t$ would also show a more pronounced centrality
dependence. However, based on the available data, this ambiguity cannot be resolved.}
\label{fig:pars}
\end{figure*}

\section{Summary}
The goal of this paper was to expand the scope of analytic relativistic hydrodynamics to higher order azimuthal symmetries,
compatible with realistic (event-by-event fluctuating) geometries. This was achieved through finding a scale variable of
suitable symmetries, through which thermodynamic quantities depend on spatial coordinates. A new exact analytic solution
of relativistic hydrodynamics was found this way, for a special class of initial conditions. Higher order flow observables
($v_n$'s) were then calculated from this model, and their model parameter dependence was investigated. It was also found
that different flow patterns may lead to the same observed $v_n$ values. Finally, we gave a set of parameters
with which our solution is found to be compatible with PHENIX data.

\section*{Acknowledgements}
M. Cs. is thankful for useful discussions with Tam\'as Cs\''org\H{o}, and acknowledges the support of the OTKA Grant No. NK101438.

\bibliography{../../../Master}

\begin{thebibliography}{25}%
\makeatletter
\providecommand \@ifxundefined [1]{%
 \@ifx{#1\undefined}
}%
\providecommand \@ifnum [1]{%
 \ifnum #1\expandafter \@firstoftwo
 \else \expandafter \@secondoftwo
 \fi
}%
\providecommand \@ifx [1]{%
 \ifx #1\expandafter \@firstoftwo
 \else \expandafter \@secondoftwo
 \fi
}%
\providecommand \natexlab [1]{#1}%
\providecommand \enquote  [1]{``#1''}%
\providecommand \bibnamefont  [1]{#1}%
\providecommand \bibfnamefont [1]{#1}%
\providecommand \citenamefont [1]{#1}%
\providecommand \href@noop [0]{\@secondoftwo}%
\providecommand \href [0]{\begingroup \@sanitize@url \@href}%
\providecommand \@href[1]{\@@startlink{#1}\@@href}%
\providecommand \@@href[1]{\endgroup#1\@@endlink}%
\providecommand \@sanitize@url [0]{\catcode `\\12\catcode `\$12\catcode
  `\&12\catcode `\#12\catcode `\^12\catcode `\_12\catcode `\%12\relax}%
\providecommand \@@startlink[1]{}%
\providecommand \@@endlink[0]{}%
\providecommand \url  [0]{\begingroup\@sanitize@url \@url }%
\providecommand \@url [1]{\endgroup\@href {#1}{\urlprefix }}%
\providecommand \urlprefix  [0]{URL }%
\providecommand \Eprint [0]{\href }%
\providecommand \doibase [0]{http://dx.doi.org/}%
\providecommand \selectlanguage [0]{\@gobble}%
\providecommand \bibinfo  [0]{\@secondoftwo}%
\providecommand \bibfield  [0]{\@secondoftwo}%
\providecommand \translation [1]{[#1]}%
\providecommand \BibitemOpen [0]{}%
\providecommand \bibitemStop [0]{}%
\providecommand \bibitemNoStop [0]{.\EOS\space}%
\providecommand \EOS [0]{\spacefactor3000\relax}%
\providecommand \BibitemShut  [1]{\csname bibitem#1\endcsname}%
\let\auto@bib@innerbib\@empty
\bibitem [{\citenamefont {Adcox}\ \emph {et~al.}(2005)\citenamefont {Adcox}
  \emph {et~al.}}]{Adcox:2004mh}%
  \BibitemOpen
  \bibfield  {author} {\bibinfo {author} {\bibfnamefont {K.}~\bibnamefont
  {Adcox}} \emph {et~al.} (\bibinfo {collaboration} {PHENIX}),\ }\href
  {\doibase 10.1016/j.nuclphysa.2005.03.086} {\bibfield  {journal} {\bibinfo
  {journal} {Nucl. Phys.}\ }\textbf {\bibinfo {volume} {A757}},\ \bibinfo
  {pages} {184} (\bibinfo {year} {2005})},\ \Eprint
  {http://arxiv.org/abs/nucl-ex/0410003} {arXiv:nucl-ex/0410003} \BibitemShut
  {NoStop}%
\bibitem [{\citenamefont {Cs\"org\H{o}}\ \emph
  {et~al.}(2004{\natexlab{a}})\citenamefont {Cs\"org\H{o}}, \citenamefont
  {Csernai}, \citenamefont {Hama},\ and\ \citenamefont
  {Kodama}}]{Csorgo:2003ry}%
  \BibitemOpen
  \bibfield  {author} {\bibinfo {author} {\bibfnamefont {T.}~\bibnamefont
  {Cs\"org\H{o}}}, \bibinfo {author} {\bibfnamefont {L.~P.}\ \bibnamefont
  {Csernai}}, \bibinfo {author} {\bibfnamefont {Y.}~\bibnamefont {Hama}}, \
  and\ \bibinfo {author} {\bibfnamefont {T.}~\bibnamefont {Kodama}},\
  }\href@noop {} {\bibfield  {journal} {\bibinfo  {journal} {Heavy Ion Phys.}\
  }\textbf {\bibinfo {volume} {A21}},\ \bibinfo {pages} {73} (\bibinfo {year}
  {2004}{\natexlab{a}})},\ \Eprint {http://arxiv.org/abs/nucl-th/0306004}
  {nucl-th/0306004} \BibitemShut {NoStop}%
\bibitem [{\citenamefont {Adare}\ \emph {et~al.}(2011)\citenamefont {Adare}
  \emph {et~al.}}]{Adare:2011tg}%
  \BibitemOpen
  \bibfield  {author} {\bibinfo {author} {\bibfnamefont {A.}~\bibnamefont
  {Adare}} \emph {et~al.} (\bibinfo {collaboration} {PHENIX Collaboration}),\
  }\href {\doibase 10.1103/PhysRevLett.107.252301} {\bibfield  {journal}
  {\bibinfo  {journal} {Phys.Rev.Lett.}\ }\textbf {\bibinfo {volume} {107}},\
  \bibinfo {pages} {252301} (\bibinfo {year} {2011})},\ \Eprint
  {http://arxiv.org/abs/1105.3928} {arXiv:1105.3928 [nucl-ex]} \BibitemShut
  {NoStop}%
\bibitem [{\citenamefont {Aamodt}\ \emph {et~al.}(2011)\citenamefont {Aamodt}
  \emph {et~al.}}]{ALICE:2011ab}%
  \BibitemOpen
  \bibfield  {author} {\bibinfo {author} {\bibfnamefont {K.}~\bibnamefont
  {Aamodt}} \emph {et~al.} (\bibinfo {collaboration} {ALICE Collaboration}),\
  }\href {\doibase 10.1103/PhysRevLett.107.032301} {\bibfield  {journal}
  {\bibinfo  {journal} {Phys.Rev.Lett.}\ }\textbf {\bibinfo {volume} {107}},\
  \bibinfo {pages} {032301} (\bibinfo {year} {2011})},\ \Eprint
  {http://arxiv.org/abs/1105.3865} {arXiv:1105.3865 [nucl-ex]} \BibitemShut
  {NoStop}%
\bibitem [{\citenamefont {Adamczyk}\ \emph {et~al.}(2013)\citenamefont
  {Adamczyk} \emph {et~al.}}]{Adamczyk:2013waa}%
  \BibitemOpen
  \bibfield  {author} {\bibinfo {author} {\bibfnamefont {L.}~\bibnamefont
  {Adamczyk}} \emph {et~al.} (\bibinfo {collaboration} {STAR Collaboration}),\
  }\href {\doibase 10.1103/PhysRevC.88.014904} {\bibfield  {journal} {\bibinfo
  {journal} {Phys.Rev.}\ }\textbf {\bibinfo {volume} {C88}},\ \bibinfo {pages}
  {014904} (\bibinfo {year} {2013})},\ \Eprint {http://arxiv.org/abs/1301.2187}
  {arXiv:1301.2187 [nucl-ex]} \BibitemShut {NoStop}%
\bibitem [{\citenamefont {Petersen}\ \emph {et~al.}(2010)\citenamefont
  {Petersen}, \citenamefont {Qin}, \citenamefont {Bass},\ and\ \citenamefont
  {Muller}}]{Petersen:2010cw}%
  \BibitemOpen
  \bibfield  {author} {\bibinfo {author} {\bibfnamefont {H.}~\bibnamefont
  {Petersen}}, \bibinfo {author} {\bibfnamefont {G.-Y.}\ \bibnamefont {Qin}},
  \bibinfo {author} {\bibfnamefont {S.~A.}\ \bibnamefont {Bass}}, \ and\
  \bibinfo {author} {\bibfnamefont {B.}~\bibnamefont {Muller}},\ }\href
  {\doibase 10.1103/PhysRevC.82.041901} {\bibfield  {journal} {\bibinfo
  {journal} {Phys.Rev.}\ }\textbf {\bibinfo {volume} {C82}},\ \bibinfo {pages}
  {041901} (\bibinfo {year} {2010})},\ \Eprint {http://arxiv.org/abs/1008.0625}
  {arXiv:1008.0625} \BibitemShut {NoStop}%
\bibitem [{\citenamefont {Schenke}\ \emph {et~al.}(2011)\citenamefont
  {Schenke}, \citenamefont {Jeon},\ and\ \citenamefont
  {Gale}}]{Schenke:2010rr}%
  \BibitemOpen
  \bibfield  {author} {\bibinfo {author} {\bibfnamefont {B.}~\bibnamefont
  {Schenke}}, \bibinfo {author} {\bibfnamefont {S.}~\bibnamefont {Jeon}}, \
  and\ \bibinfo {author} {\bibfnamefont {C.}~\bibnamefont {Gale}},\ }\href
  {\doibase 10.1103/PhysRevLett.106.042301} {\bibfield  {journal} {\bibinfo
  {journal} {Phys.Rev.Lett.}\ }\textbf {\bibinfo {volume} {106}},\ \bibinfo
  {pages} {042301} (\bibinfo {year} {2011})},\ \Eprint
  {http://arxiv.org/abs/1009.3244} {arXiv:1009.3244 [hep-ph]} \BibitemShut
  {NoStop}%
\bibitem [{\citenamefont {Bravina}\ \emph {et~al.}(2014)\citenamefont
  {Bravina}, \citenamefont {Johansson}, \citenamefont {Eyyubova}, \citenamefont
  {Korotkikh}, \citenamefont {Lokhtin} \emph {et~al.}}]{Bravina:2013xla}%
  \BibitemOpen
  \bibfield  {author} {\bibinfo {author} {\bibfnamefont {L.}~\bibnamefont
  {Bravina}}, \emph
  {et~al.},\ }\href {\doibase 10.1140/epjc/s10052-014-2807-5} {\bibfield
  {journal} {\bibinfo  {journal} {Eur.Phys.J.}\ }\textbf {\bibinfo {volume}
  {C74}},\ \bibinfo {pages} {2807} (\bibinfo {year} {2014})},\ \Eprint
  {http://arxiv.org/abs/1311.7054} {arXiv:1311.7054 [nucl-th]} \BibitemShut
  {NoStop}%
\bibitem [{\citenamefont {Chatterjee}\ \emph {et~al.}(2014)\citenamefont
  {Chatterjee}, \citenamefont {Srivastava},\ and\ \citenamefont
  {Renk}}]{Chatterjee:2014nta}%
  \BibitemOpen
  \bibfield  {author} {\bibinfo {author} {\bibfnamefont {R.}~\bibnamefont
  {Chatterjee}}, \bibinfo {author} {\bibfnamefont {D.~K.}\ \bibnamefont
  {Srivastava}}, \ and\ \bibinfo {author} {\bibfnamefont {T.}~\bibnamefont
  {Renk}},\ }\href@noop {} {\  (\bibinfo {year} {2014})},\ \Eprint
  {http://arxiv.org/abs/1401.7464} {arXiv:1401.7464 [hep-ph]} \BibitemShut
  {NoStop}%
\bibitem [{\citenamefont {Csan\'ad}\ \emph {et~al.}(2012)\citenamefont
  {Csan\'ad}, \citenamefont {Nagy},\ and\ \citenamefont
  {L\"ok\"os}}]{Csanad:2012hr}%
  \BibitemOpen
  \bibfield  {author} {\bibinfo {author} {\bibfnamefont {M.}~\bibnamefont
  {Csan\'ad}}, \bibinfo {author} {\bibfnamefont {M.}~\bibnamefont {Nagy}}, \
  and\ \bibinfo {author} {\bibfnamefont {S.}~\bibnamefont {L\"ok\"os}},\ }\href
  {\doibase 10.1140/epja/i2012-12173-7} {\bibfield  {journal} {\bibinfo
  {journal} {Eur.Phys.J.}\ }\textbf {\bibinfo {volume} {A48}},\ \bibinfo
  {pages} {173} (\bibinfo {year} {2012})},\ \Eprint
  {http://arxiv.org/abs/1205.5965} {arXiv:1205.5965 [nucl-th]} \BibitemShut
  {NoStop}%
\bibitem [{\citenamefont {Csan\'ad}(2009)}]{Csanad:2009sk}%
  \BibitemOpen
  \bibfield  {author} {\bibinfo {author} {\bibfnamefont {M.}~\bibnamefont
  {Csan\'ad}},\ }\href@noop {} {\bibfield  {journal} {\bibinfo  {journal} {Acta
  Phys. Polon.}\ }\textbf {\bibinfo {volume} {B40}},\ \bibinfo {pages} {1193}
  (\bibinfo {year} {2009})},\ \Eprint {http://arxiv.org/abs/0903.1278}
  {arXiv:0903.1278 [nucl-th]} \BibitemShut {NoStop}%
\bibitem [{\citenamefont {Landau}(1953)}]{Landau:1953gs}%
  \BibitemOpen
  \bibfield  {author} {\bibinfo {author} {\bibfnamefont {L.~D.}\ \bibnamefont
  {Landau}},\ }\href@noop {} {\bibfield  {journal} {\bibinfo  {journal} {Izv.
  Akad. Nauk SSSR Ser. Fiz.}\ }\textbf {\bibinfo {volume} {17}},\ \bibinfo
  {pages} {51} (\bibinfo {year} {1953})}\BibitemShut {NoStop}%
\bibitem [{\citenamefont {Belenkij}\ and\ \citenamefont
  {Landau}(1956)}]{Belenkij:1956cd}%
  \BibitemOpen
  \bibfield  {author} {\bibinfo {author} {\bibfnamefont {S.~Z.}\ \bibnamefont
  {Belenkij}}\ and\ \bibinfo {author} {\bibfnamefont {L.~D.}\ \bibnamefont
  {Landau}},\ }\href@noop {} {\bibfield  {journal} {\bibinfo  {journal} {Nuovo
  Cim. Suppl.}\ }\textbf {\bibinfo {volume} {3S10}},\ \bibinfo {pages} {15}
  (\bibinfo {year} {1956})}\BibitemShut {NoStop}%
\bibitem [{\citenamefont {Hwa}(1974)}]{Hwa:1974gn}%
  \BibitemOpen
  \bibfield  {author} {\bibinfo {author} {\bibfnamefont {R.~C.}\ \bibnamefont
  {Hwa}},\ }\href@noop {} {\bibfield  {journal} {\bibinfo  {journal} {Phys.
  Rev.}\ }\textbf {\bibinfo {volume} {D10}},\ \bibinfo {pages} {2260} (\bibinfo
  {year} {1974})}\BibitemShut {NoStop}%
\bibitem [{\citenamefont {Bjorken}(1983)}]{Bjorken:1982qr}%
  \BibitemOpen
  \bibfield  {author} {\bibinfo {author} {\bibfnamefont {J.~D.}\ \bibnamefont
  {Bjorken}},\ }\href@noop {} {\bibfield  {journal} {\bibinfo  {journal} {Phys.
  Rev.}\ }\textbf {\bibinfo {volume} {D27}},\ \bibinfo {pages} {140} (\bibinfo
  {year} {1983})}\BibitemShut {NoStop}%
\bibitem [{\citenamefont {Bialas}\ \emph {et~al.}(2007)\citenamefont {Bialas},
  \citenamefont {Janik},\ and\ \citenamefont {Peschanski}}]{Bialas:2007iu}%
  \BibitemOpen
  \bibfield  {author} {\bibinfo {author} {\bibfnamefont {A.}~\bibnamefont
  {Bialas}}, \bibinfo {author} {\bibfnamefont {R.~A.}\ \bibnamefont {Janik}}, \
  and\ \bibinfo {author} {\bibfnamefont {R.~B.}\ \bibnamefont {Peschanski}},\
  }\href {\doibase 10.1103/PhysRevC.76.054901} {\bibfield  {journal} {\bibinfo
  {journal} {Phys. Rev.}\ }\textbf {\bibinfo {volume} {C76}},\ \bibinfo {pages}
  {054901} (\bibinfo {year} {2007})},\ \Eprint {http://arxiv.org/abs/0706.2108}
  {arXiv:0706.2108 [nucl-th]} \BibitemShut {NoStop}%
\bibitem [{\citenamefont {Beuf}\ \emph {et~al.}(2008)\citenamefont {Beuf},
  \citenamefont {Peschanski},\ and\ \citenamefont {Saridakis}}]{Beuf:2008vd}%
  \BibitemOpen
  \bibfield  {author} {\bibinfo {author} {\bibfnamefont {G.}~\bibnamefont
  {Beuf}}, \bibinfo {author} {\bibfnamefont {R.}~\bibnamefont {Peschanski}}, \
  and\ \bibinfo {author} {\bibfnamefont {E.~N.}\ \bibnamefont {Saridakis}},\
  }\href {\doibase 10.1103/PhysRevC.78.064909} {\bibfield  {journal} {\bibinfo
  {journal} {Phys.Rev.}\ }\textbf {\bibinfo {volume} {C78}},\ \bibinfo {pages}
  {064909} (\bibinfo {year} {2008})},\ \Eprint {http://arxiv.org/abs/0808.1073}
  {arXiv:0808.1073 [nucl-th]} \BibitemShut {NoStop}%
\bibitem [{\citenamefont {Peschanski}\ and\ \citenamefont
  {Saridakis}(2011)}]{Peschanski:2010cs}%
  \BibitemOpen
  \bibfield  {author} {\bibinfo {author} {\bibfnamefont {R.}~\bibnamefont
  {Peschanski}}\ and\ \bibinfo {author} {\bibfnamefont {E.~N.}\ \bibnamefont
  {Saridakis}},\ }\href {\doibase 10.1016/j.nuclphysa.2010.10.009} {\bibfield
  {journal} {\bibinfo  {journal} {Nucl.Phys.}\ }\textbf {\bibinfo {volume}
  {A849}},\ \bibinfo {pages} {147} (\bibinfo {year} {2011})},\ \Eprint
  {http://arxiv.org/abs/1006.1603} {arXiv:1006.1603 [hep-th]} \BibitemShut
  {NoStop}%
\bibitem [{\citenamefont {Cs\"org\H{o}}\ \emph
  {et~al.}(2004{\natexlab{b}})\citenamefont {Cs\"org\H{o}}, \citenamefont
  {Grassi}, \citenamefont {Hama},\ and\ \citenamefont
  {Kodama}}]{Csorgo:2002ki}%
  \BibitemOpen
  \bibfield  {author} {\bibinfo {author} {\bibfnamefont {T.}~\bibnamefont
  {Cs\"org\H{o}}}, \bibinfo {author} {\bibfnamefont {F.}~\bibnamefont
  {Grassi}}, \bibinfo {author} {\bibfnamefont {Y.}~\bibnamefont {Hama}}, \ and\
  \bibinfo {author} {\bibfnamefont {T.}~\bibnamefont {Kodama}},\ }\href@noop {}
  {\bibfield  {journal} {\bibinfo  {journal} {Heavy Ion Physics}\ }\textbf
  {\bibinfo {volume} {A21}},\ \bibinfo {pages} {53} (\bibinfo {year}
  {2004}{\natexlab{b}})},\ \Eprint {http://arxiv.org/abs/hep-ph/0203204}
  {hep-ph/0203204} \BibitemShut {NoStop}%
\bibitem [{\citenamefont {Cs\"org\H{o}}\ \emph {et~al.}(2003)\citenamefont
  {Cs\"org\H{o}}, \citenamefont {Grassi}, \citenamefont {Hama},\ and\
  \citenamefont {Kodama}}]{Csorgo:2003rt}%
  \BibitemOpen
  \bibfield  {author} {\bibinfo {author} {\bibfnamefont {T.}~\bibnamefont
  {Cs\"org\H{o}}}, \bibinfo {author} {\bibfnamefont {F.}~\bibnamefont
  {Grassi}}, \bibinfo {author} {\bibfnamefont {Y.}~\bibnamefont {Hama}}, \ and\
  \bibinfo {author} {\bibfnamefont {T.}~\bibnamefont {Kodama}},\ }\href@noop {}
  {\bibfield  {journal} {\bibinfo  {journal} {Phys. Lett.}\ }\textbf {\bibinfo
  {volume} {B565}},\ \bibinfo {pages} {107} (\bibinfo {year} {2003})},\ \Eprint
  {http://arxiv.org/abs/nucl-th/0305059} {nucl-th/0305059} \BibitemShut
  {NoStop}%
\bibitem [{\citenamefont {Liao}\ and\ \citenamefont
  {Koch}(2009)}]{Liao:2009zg}%
  \BibitemOpen
  \bibfield  {author} {\bibinfo {author} {\bibfnamefont {J.}~\bibnamefont
  {Liao}}\ and\ \bibinfo {author} {\bibfnamefont {V.}~\bibnamefont {Koch}},\
  }\href {\doibase 10.1103/PhysRevC.80.034904} {\bibfield  {journal} {\bibinfo
  {journal} {Phys.Rev.}\ }\textbf {\bibinfo {volume} {C80}},\ \bibinfo {pages}
  {034904} (\bibinfo {year} {2009})},\ \Eprint {http://arxiv.org/abs/0905.3406}
  {arXiv:0905.3406 [nucl-th]} \BibitemShut {NoStop}%
\bibitem [{\citenamefont {Lin}\ and\ \citenamefont {Liao}(2010)}]{Lin:2009kv}%
  \BibitemOpen
  \bibfield  {author} {\bibinfo {author} {\bibfnamefont {S.}~\bibnamefont
  {Lin}}\ and\ \bibinfo {author} {\bibfnamefont {J.}~\bibnamefont {Liao}},\
  }\href {\doibase 10.1016/j.nuclphysa.2010.02.011} {\bibfield  {journal}
  {\bibinfo  {journal} {Nucl.Phys.}\ }\textbf {\bibinfo {volume} {A837}},\
  \bibinfo {pages} {195} (\bibinfo {year} {2010})},\ \Eprint
  {http://arxiv.org/abs/0909.2284} {arXiv:0909.2284 [nucl-th]} \BibitemShut
  {NoStop}%
\bibitem [{\citenamefont {Gubser}(2010)}]{Gubser:2010ze}%
  \BibitemOpen
  \bibfield  {author} {\bibinfo {author} {\bibfnamefont {S.~S.}\ \bibnamefont
  {Gubser}},\ }\href {\doibase 10.1103/PhysRevD.82.085027} {\bibfield
  {journal} {\bibinfo  {journal} {Phys.Rev.}\ }\textbf {\bibinfo {volume}
  {D82}},\ \bibinfo {pages} {085027} (\bibinfo {year} {2010})},\ \Eprint
  {http://arxiv.org/abs/1006.0006} {arXiv:1006.0006 [hep-th]} \BibitemShut
  {NoStop}%
\bibitem [{\citenamefont {Hatta}\ \emph {et~al.}(2014)\citenamefont {Hatta},
  \citenamefont {Noronha},\ and\ \citenamefont {Xiao}}]{Hatta:2014gqa}%
  \BibitemOpen
  \bibfield  {author} {\bibinfo {author} {\bibfnamefont {Y.}~\bibnamefont
  {Hatta}}, \bibinfo {author} {\bibfnamefont {J.}~\bibnamefont {Noronha}}, \
  and\ \bibinfo {author} {\bibfnamefont {B.-W.}\ \bibnamefont {Xiao}},\ }\href
  {\doibase 10.1103/PhysRevD.89.051702} {\bibfield  {journal} {\bibinfo
  {journal} {Phys.Rev.}\ }\textbf {\bibinfo {volume} {D89}},\ \bibinfo {pages}
  {051702} (\bibinfo {year} {2014})},\ \Eprint {http://arxiv.org/abs/1401.6248}
  {arXiv:1401.6248 [hep-th]} \BibitemShut {NoStop}%
\bibitem [{\citenamefont {Csan\'ad}\ and\ \citenamefont
  {Vargyas}(2010)}]{Csanad:2009wc}%
  \BibitemOpen
  \bibfield  {author} {\bibinfo {author} {\bibfnamefont {M.}~\bibnamefont
  {Csan\'ad}}\ and\ \bibinfo {author} {\bibfnamefont {M.}~\bibnamefont
  {Vargyas}},\ }\href {\doibase 10.1140/epja/i2010-10973-3} {\bibfield
  {journal} {\bibinfo  {journal} {Eur. Phys. J.}\ }\textbf {\bibinfo {volume}
  {A44}},\ \bibinfo {pages} {473} (\bibinfo {year} {2010})},\ \Eprint
  {http://arxiv.org/abs/0909.4842} {arXiv:0909.4842 [nucl-th]} \BibitemShut
  {NoStop}%
\end{thebibliography}%

\end{document}